\documentclass[11pt]{article}
\pdfoutput=1

\usepackage{euscript}
\usepackage{amssymb}
\usepackage{amsfonts}
\usepackage{amsbsy}
\usepackage{epsfig}
\usepackage{amsthm}
\usepackage{amscd}
\usepackage{amstext}
\usepackage{verbatim}
\usepackage{amsmath}
\usepackage{cancel}
\usepackage{capt-of}
\usepackage{empheq}
\usepackage{subfigure}
\usepackage{xcolor}

\usepackage{cite}
\usepackage{bm}
\usepackage{authblk}
\usepackage[T1]{fontenc}
\usepackage{braket}

\usepackage[pdftex,linktoc=all]{hyperref}
\hypersetup{ 
colorlinks=true, 
linkcolor=black, 
citecolor=red, 
}

\def\ben{\begin{equation}}
\def\een{\end{equation}}
\def\half{{\textstyle{\frac{1}{2}}}}

\let\pa=\partial
\def\be{\begin{equation}}
\def\ee{\end{equation}}
\def\beq{\begin{equation}}
\def\eeq{\end{equation}}
\def\ba{\begin{array}}
\def\ea{\end{array}}

\def\dalemb#1#2{{\vbox{\hrule height .#2pt
       \hbox{\vrule width.#2pt height#1pt \kern#1pt
               \vrule width.#2pt}
       \hrule height.#2pt}}}

\newcommand{\bea}{\begin{eqnarray}}
\newcommand{\eea}{\end{eqnarray}}

\def\vep{{\varepsilon}}

\makeatletter
\newcommand*\bigcdot{\mathpalette\bigcdot@{.5}}
\newcommand*\bigcdot@[2]{\mathbin{\vcenter{\hbox{\scalebox{#2}{$\m@th#1\bullet$}}}}}
\makeatother

\renewcommand{\eqref}[1]{(\ref{#1})}

\def\Lag{{\mathcal{L}}}

\def\ocal{{\mathcal{O}}}

\title{Holographic flows from CFT to the Kasner universe}
\author{Alexander Frenkel, Sean A. Hartnoll, Jorrit Kruthoff, Zhengyan D. Shi}
\affil{Department of Physics, Stanford University, \\
Stanford, CA 94305-4060, USA}
\date{}                     

\begin{document}
\frenchspacing

\maketitle

\begin{abstract}

The Schwarzschild singularity is known to be classically unstable.
We demonstrate a simple holographic consequence of this fact, focusing on a perturbation that is uniform in boundary space and time. Deformation of the thermal state of the dual CFT by a relevant operator triggers a nonzero temperature holographic renormalization group flow in the bulk. This flow continues smoothly through the horizon and, at late interior time, deforms the Schwarzschild singularity into a more general Kasner universe. We show that the deformed near-singularity, trans-horizon Kasner exponents determine specific non-analytic corrections to the thermal correlation functions of heavy operators in the dual CFT, in the analytically continued `near-singularity' regime.
\end{abstract}

\newpage

\tableofcontents

\section{Introduction}

A longstanding promise of holographic duality is to shed light upon the black hole interior. Holographic probes of the black hole interior include analytically continued correlation functions \cite{Fidkowski:2003nf, Festuccia:2005pi}, entanglement entropy \cite{Hartman:2013qma} and perhaps complexity \cite{Stanford:2014jda, Brown:2015bva}. The most dramatic aspect of the black hole interior is the inevitability of a spacetime singularities \cite{Hawking:1969sw}. Spacelike singularities, at which time `ends' are the most conceptually challenging in this regard and also bring out the strong similarity between black hole interiors and cosmology.

The most familiar black hole interior in holography is that of the eternal Schwarzschild-AdS black hole. This black hole plays an important role in describing the thermofield double state of the dual CFT \cite{Maldacena:2001kr}. However, while the exterior geometry of these black holes is dynamically stable (for common choices of matter content), the singularity is not. It has been known for some time that e.g. scalar fields blow up upon approach to the singularity \cite{Doroshkevich:1978aq,Fournodavlos:2018lrk}. More generally it is known that the Schwarzschild singularity is very finely tuned within the space of possible late time behaviors of gravity and therefore cannot be a generic late time solution, e.g. \cite{BKL}. Clearly, the instability of the Schwarzschild singularity should be taken into account when holographic probes of the black hole interior are considered.

The generic late-time behavior inside the horizon is expected to be highly inhomogeneous. However, even restricting to geometries that retain the spacetime symmetries of Schwarzschild-AdS, the Schwarzschild singularity is fine-tuned. The main purpose of our work is to exhibit some consequences of this fact. We will focus on four dimensional planar AdS black holes, and will couple gravity to a scalar field (this will be important to retain boundary spatial isotropy). The Lagrangian density is
\be\label{eq:action}
\Lag = \frac{1}{2 \kappa^2} \left( R + 6 \right) - \frac{1}{2} \left(g^{ab} \pa_a \phi \pa_b \phi - m^2 \phi^2\right) \,.
\ee
We have set the AdS radius to one. The black holes we consider will have the form
\be\label{eq:adsBH}
ds^2 = \frac{1}{r^2} \left( - f(r) e^{-\chi(r)} dt^2 + \frac{dr^2}{f(r)} + dx^2 + dy^2 \right) \,,
\ee
with the scalar field $\phi = \frac{1}{\sqrt{2} \kappa} \phi(r)$.
In our conventions the radial coordinate $r \to 0$ at the AdS boundary and $r \to \infty$ at the singularity. The horizon is at $f(r_+) = 0$. The planar Schwarzschild-AdS solution has $\chi = 0$ and $f = 1 - (r/r_+)^3$. 
The near-singularity behavior of the general class of spacetimes (\ref{eq:adsBH}) has the Kasner universe form \cite{Kasner, Belinski:1973zz}
\be\label{eq:kasner}
ds^2 \sim - d\tau^2 + \tau^{2 p_t} dt^2 + \tau^{2 p_x} \left(dx^2 + dy^2 \right) \,, \qquad \phi(r) \sim - \sqrt{2} p_\phi \log \tau \,.
\ee
Here $\tau$ is obtained from the radial coordinate $r$, which is timelike inside the horizon. The Kasner exponents obey $p_t + 2 p_x = 1$ and $p_\phi^2 + p_t^2 + 2 p_x^2 = 1$.
The Schwarzschild singularity has $p_t = - \frac{1}{3}, p_x = \frac{2}{3}, p_\phi = 0$. Without the scalar field, this would be the only nontrivial solution that is isotropic in $x$ and $y$. With the scalar field, however, the Schwarzschild singularity lies within a one-parameter family of $x-y$ isotropic near-singularity behaviors.

\begin{figure}[ht!]
    \vskip 0.2in
    \centering
    \includegraphics[width=0.35\textwidth]{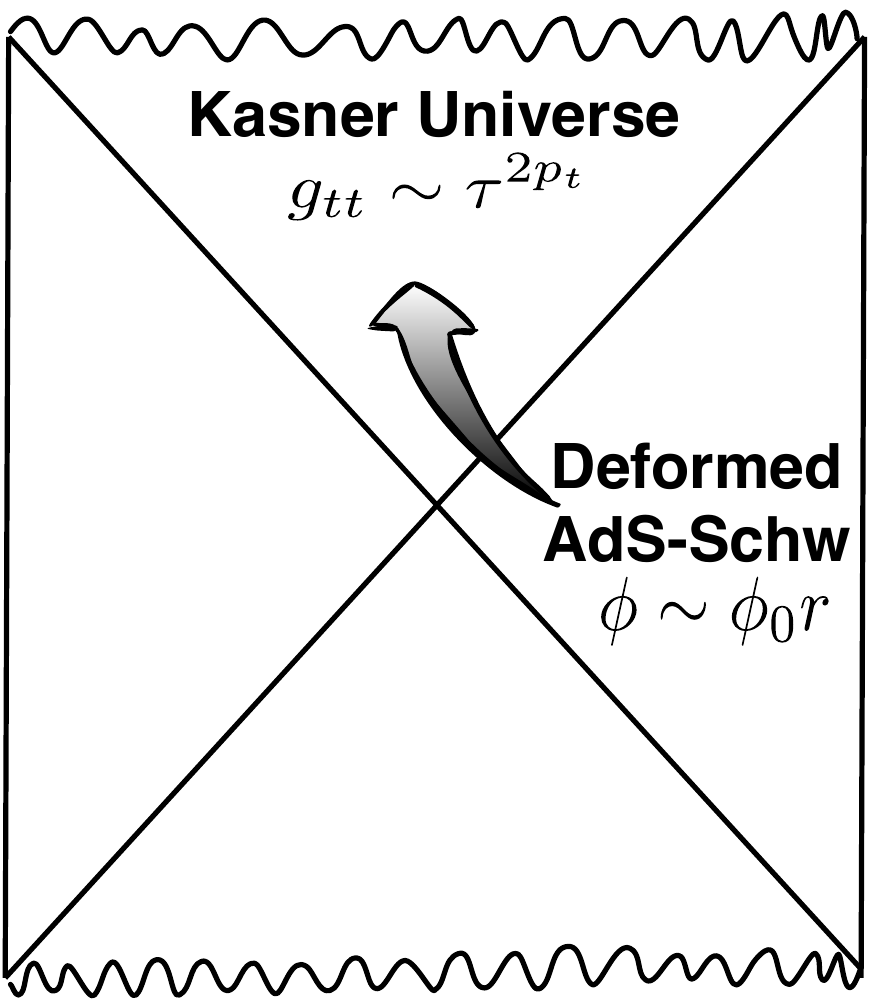}
    \caption{The AdS-Schwarzschild solution at temperature $T$ is deformed at both AdS boundaries ($r \to 0$) by a relevant operator $\ocal$ with coupling $\phi_0$. We take $\ocal$ to have dimension $\Delta = 2$ in 2+1 boundary dimensions ($t,x,y$). The deformation is uniform in $t,x,y$. The deformed solution extends smoothly through the horizon and tends towards a Kasner universe near the singularity $\tau \to 0$. Here $\tau(r)$ is the proper time to the singularity. The Kasner scaling exponent $p_t$ is determined by the dimensionless ratio $\phi_0/T$.}
    \label{fig:kasnerflow}
    \vskip -0.1in
\end{figure}

Using the theory (\ref{eq:action}) we will show that if the thermal state of the dual CFT is perturbed by a relevant operator, sourcing the field $\phi$ at the AdS boundary, then the near-singularity Kasner scaling exponents are shifted away from their Schwarzschild value. This amounts to a generalization of the notion of a holographic renormalization group flow. Usually these are zero temperature solutions that interpolate from a UV to an IR radial scaling fixed point, e.g. \cite{Skenderis:1999mm}. Aspects of nonzero temperature flows outside the horizon have been considered in \cite{Gursoy:2018umf}. Here we are discussing flows in thermal states that interpolate from a UV radial scaling to a timelike scaling towards a late time singularity in the black hole interior. The Kasner exponents play a role analogous to the scaling dimensions of operators in the CFT. This is illustrated in the Fig. \ref{fig:kasnerflow}.

Previous discussions of Kasner universes and associated singularities in a holographic context have considered time-dependent CFT processes, in which the CFT itself is placed on a background conformal to the Kasner universe \cite{Das:2006dz, Engelhardt:2013jda, Engelhardt:2014mea, Engelhardt:2015gta, Barbon:2015ria}. In our setup the cosmological time dependence instead emerges dynamically in the black hole interior. The CFT is in a relatively mundane time-independent thermal state, deformed by a relevant operator.

Black hole singularities can be probed from the dual CFT using spacelike geodesics that go from one boundary to the other, traversing the Einstein-Rosen bridge and coming close to the singularity \cite{Fidkowski:2003nf}. As the boundary time $t$ approaches a specific value $t_\text{sing}$, the geodesic bounces off the singularity and the regularized length $L$ of the geodesic diverges $L \sim 2 \log(2 |t - t_\text{sing}|)$. This geodesic length contributes to a Schwinger-Keldysh correlation function of heavy operators in the dual CFT. While it is not the dominant saddle of the correlation function in the physical regime, it can be accessed via analytic continuation \cite{Fidkowski:2003nf, Festuccia:2005pi}.

We show that the Kasner exponent $p_t$ determines specific non-analytic corrections to the correlation function in the regime described by the bouncing geodesic. Firstly, the length of the geodesic receives a correction $L = \cdots + |t - t_\text{sing}|^{-1/p_t} + \cdots$. This result is given explicitly in (\ref{eq:Lt}) below. For the Schwarzschild singularity $p_t = -1/3$ and hence this correction is analytic in that case. However, it is non-analytic for general Kasner universes. Contributions to $L$ arising from the intermediate geometry are always analytic, while contributions from the near-boundary regime are determined by the UV scaling exponent $\Delta$ of the relevant operator. This non-analytic correction given above is therefore a well-defined signature of the singularity. Secondly, we show that corrections to the correlation function due to a finite mass $M$ of the heavy probe operator become large in the regime $M |t - t_\text{sing}|^{-1/p_t} \lesssim 1$. Note that $p_t < 0$. The explicit result is in (\ref{eq:theanswer}) below. The Kasner exponent therefore controls how sensitive a large but finite mass operator is to the singular regime.

While our corrections are subleading compared to the dominant $L \sim 2 \log(2 |t - t_\text{sing}|)$ behavior, they are unversival in the sense of isolating a contribution entirely from the near-singularity spacetime. In contrast, $t_\text{sing}$ is an integral over a null geodesic from the boundary to the singularity and the logarithmic behavior of $L$ originates from near the AdS boundary.

Many of the deeper questions in classical general relativity involving the instability (or not) of singularities are concerned with spatial inhomogeneity, chaos and with the interior of charged and rotating black holes. We shall not touch on those questions here.
Holographic work on the stability of the interior of charged and rotating horizons can be found in the recent papers \cite{Dias:2019ery,Balasubramanian:2019qwk} and references therein. 
We hope that our results here can be a first step in moving beyond the non-generic and classically unstable black hole interiors that have been the focus of most previous holographic work.

\section{Thermal holographic flows from AdS to Kasner}

We can now construct explicit examples of holographic flows from the AdS boundary to a Kasner singularity inside a black hole horizon. We will find numerical solutions to the Einstein-scalar theory (\ref{eq:action}) of the form (\ref{eq:adsBH}). In the `infalling' coordinates
\be
ds^2 = \frac{1}{r^2} \left( - f(r) e^{-\chi(r)} du^2 + 2 e^{-\chi(r)/2} du dr + dx^2 + dy^2  \right) \,,
\ee
the metric is regular at the horizon where $f(r_+) = 0$. Recall that $\pa_r$ is spacelike for $f > 0$ and timelike for $f < 0$. The AdS boundary is at $r=0$ and the singularity will be at $r \to \infty$.

We will focus on the conformally coupled case with $m^2 = - 2$ (this negative mass squared is above the Breitenlohner-Freedman bound). The precise value of the mass is not important, but it should correspond to a relevant deformation of the boundary CFT. The
Einstein-scalar equations of motion are solved by radial functions obeying
\bea
\phi'' + \left(\frac{f'}{f} - \frac{2}{r} - \frac{\chi'}{2} \right)\phi' + \frac{2}{r^2 f} \phi & = & 0 \,, \\
\chi' - \frac{2 f'}{f} - \frac{\phi^2}{r f} - \frac{6}{r f} + \frac{6}{r} & = & 0 \,, \\
\chi' - \frac{r}{2} (\phi')^2 & = & 0 \,.
\eea
And the general near-boundary behavior, as $r \to 0$, is
\begin{align}
\phi = \phi_o  r + \langle \ocal \rangle r^2 + \cdots \,, \; \quad
\chi  = \frac{\phi_o ^2}{4} r^2 + \frac{2 \phi_o \langle \ocal \rangle}{3} r^3 + \cdots \,, \; \quad e^{-\chi} f =  1 - \braket{T_{tt}} r^3 + \cdots \,. \label{eq:nearb}
\end{align}
Here $\phi_0$ is the source for the dual boundary operator, with expectation value $\langle \ocal \rangle$. We are using `standard' quantization of the scalar field, wherein $\ocal$ has dimension $\Delta = 2$. We have chosen the normalization of time at the boundary so that $\chi \to 0$ as $r \to 0$. At the horizon $\chi \to \chi_+$ while $f$ vanishes. With this normalization, the temperature of the state is
\be\label{eq:temp}
T = \frac{|f_+'| e^{- \chi_+/2}}{4 \pi} \,.
\ee
Here $f'_+ = f'(r_+)$ and $\chi_+ = \chi(r_+)$. Finally, $\braket{T_{tt}}$ is the energy density of the thermal state.

\begin{figure}[ht!]
    \vskip 0.1in
    \centering
    \includegraphics[width=0.75\textwidth]{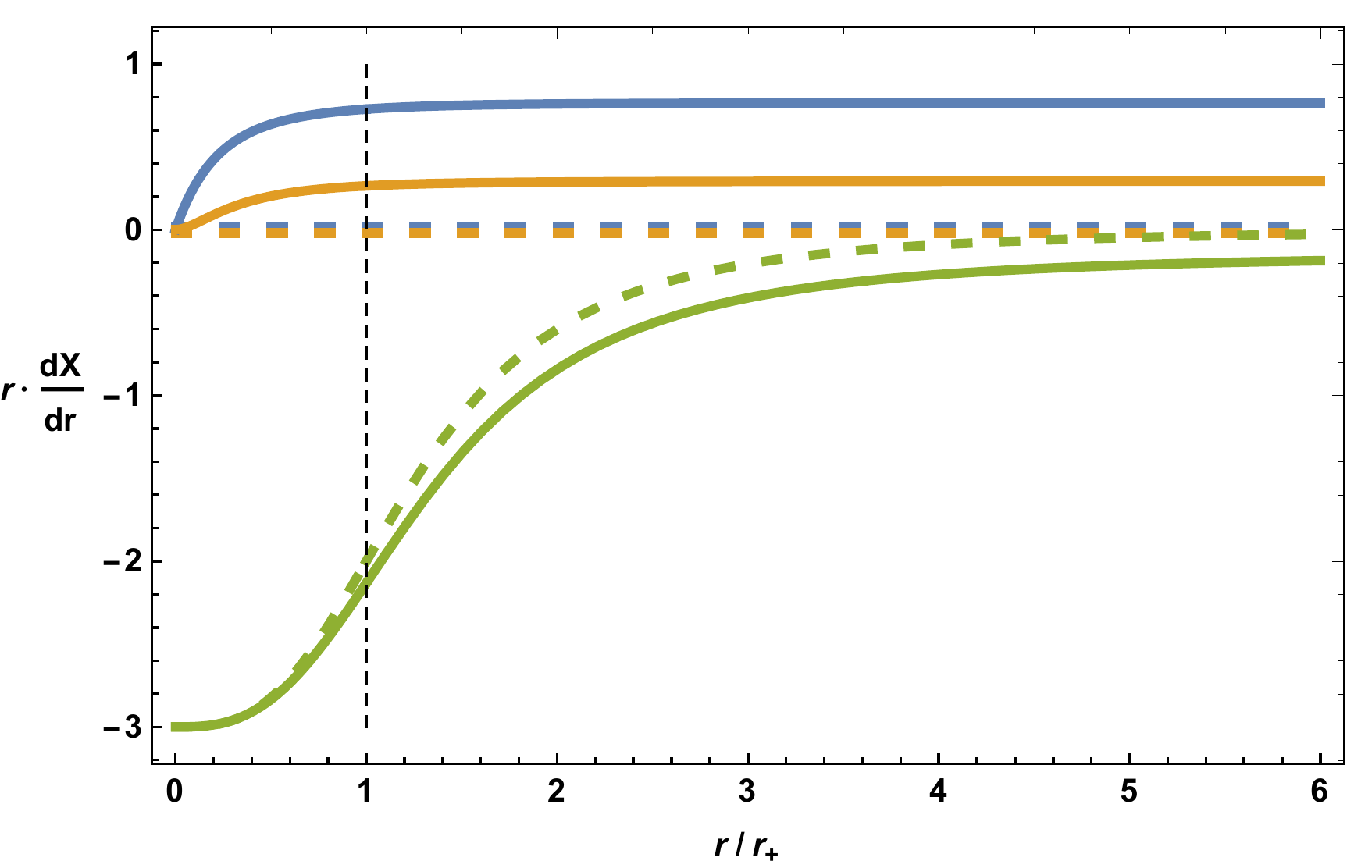}
    \caption{Flow from AdS (at $r=0$) to a Kasner cosmology (as $r\to\infty$). From top to bottom $X = \phi(r)$, $X = \chi(r)$ and $X = \log g_{tt}'(r)$. For all of these quantities $r\, dX/dr$ goes to zero at the Schwarzschild singularity (dashed curves) but tend to a constant at more general Kasner singularities (solid curves). These constants are determined by the Kasner exponents. The solid curves shown correspond to a particular flow generated by $\phi_o/T = 12.25$. The vertical dashed black line shows the location of the horizon.}
    \label{fig:flow}
    \vskip -0.1in
\end{figure}

Imposing regularity at the horizon, $r = r_+$, fixes $\braket{T_{tt}}$ and $\langle \ocal \rangle$ in terms of $T$ and $\phi_o$. Physical solutions are therefore labelled by a single dimensionless parameter $\phi_o/T$. 
Integration from the boundary, through the horizon and to the singularity will therefore determine the near-singularity behavior in terms of the ratio $\phi_o/T$. An example of such a solution is shown in Fig. \ref{fig:flow}. Such solutions are found numerically using the methods described in e.g. \cite{Hartnoll:2008kx}. The only difference is that one integrates from the horizon both to the AdS boundary and towards the singularity.

The near singularity large $r$ scaling behavior seen in Fig. \ref{fig:flow} can be understood from the general asymptotic behavior of solutions. As $r \to \infty$, the equations of motion imply that
\be\label{eq:cs}
\phi = 2 c \, \log r + \cdots \,, \qquad
\chi = 2 c^2 \log r + \chi_1 + \cdots \,, \qquad
f = - f_1 r^{3 + c^2} + \cdots \,. 
\ee
Here $c$ is a constant, with $c=0$ for Schwarzschild.
At $\ocal(c)$ this is the logarithmic growth of a (spatially uniform) scalar field towards the Schwarzschild singularity described in \cite{Doroshkevich:1978aq,Fournodavlos:2018lrk}. The $\ocal(c^2)$ terms describe the backreaction of this instability on the metric to a new self-consistent scaling form at late interior time. Indeed, setting the timelike coordinate $r^{(3 + c^2)} = 1/\tau^2$, the spacetime near the singularity approaches the Kasner form (\ref{eq:kasner}), with exponents
\be\label{eq:ppp}
p_x = p_y = \frac{2}{3 + c^2} \,, \quad p_t = \frac{c^2 - 1}{3 + c^2} \,, \quad p_\phi = \frac{2 \sqrt{2} c}{3+ c^2} \,.
\ee

\begin{figure}[ht!]
    \vskip 0.1in
    \centering
    \includegraphics[width=0.75\textwidth]{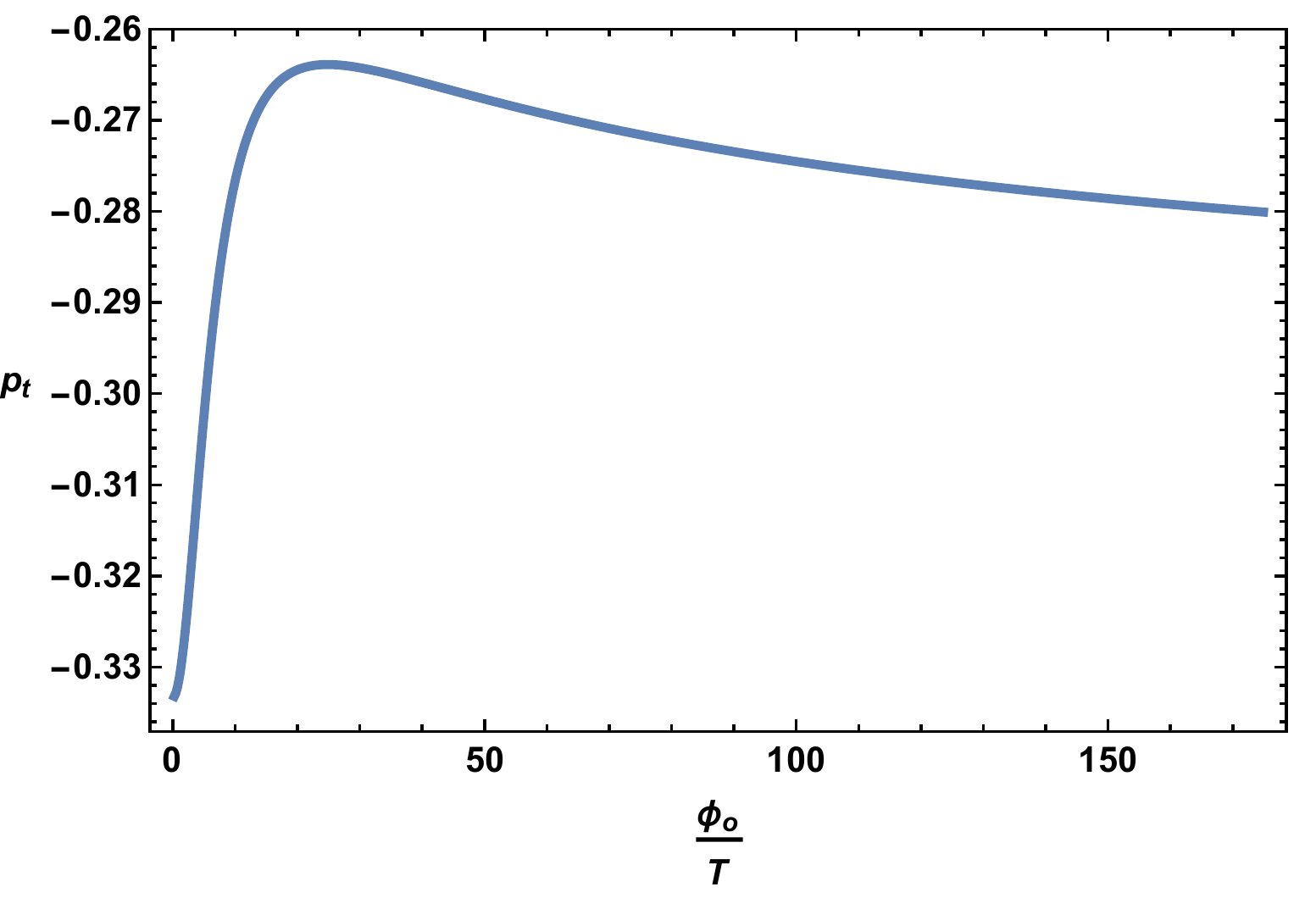}
    \vskip -0.1in
    \caption{Emergent Kasner exponent $p_t$ as a function of the dimensionless CFT deformation $\phi_o/T$. The exponents $p_x = p_y$ and $p_\phi$ are fixed by the relations below equation (\ref{eq:kasner}). As $\phi_o/T \to \infty$ we believe that $p_t$ returns to the Schwarzschild singularity value of $-1/3$. While the numerics outside the horizon are delicate in this limit, in the interior $c$ is seen to decrease to zero as the value of the scalar field on the horizon becomes large.}
    \label{fig:kasner}
    \vskip -0.1in
\end{figure}

The large $r$ behavior in Fig. \ref{fig:flow} is therefore controlled by the Kasner exponents. Fig. \ref{fig:flow} shows a holographic flow from an AdS boundary to an interior Kasner cosmology. For every choice of the dimensionless CFT parameter $\phi_o/T$, we obtain an emergent Kasner scaling determined by the exponent $p_t$. This relationship is shown in Fig. \ref{fig:kasner}.

Fig. \ref{fig:kasner} demonstrates the anticipated fine-tuned nature
of the interior Schwarzschild singularity (with $p_t = -1/3$). A deformation of the exterior that preserves the spacetime symmetries of the thermal CFT state changes the near-singularity scaling exponents. 
This can be thought of as a dynamical instability of the Schwarzschild singularity at late interior time $r$. We now describe how to extract these exponents using boundary probes.

\section{Probes of the Kasner exponent}

\subsection{Non-analyticities in the geodesic length}

Spacelike geodesics can cross the Einstein-Rosen bridge from one side of the black hole to the other. In the limit that the geodesics become almost null, they probe the vicinity of the interior singularity \cite{Fidkowski:2003nf}. These geodesics contribute to an analytic continuation of the Green's function of a large dimension operator in the dual CFT \cite{Fidkowski:2003nf, Festuccia:2005pi}. In this section we adapt the analysis in \cite{Fidkowski:2003nf} to the case of AdS to Kasner flows. The result of this section is equation (\ref{eq:Lt}), showing that the near-singularity Kasner exponent $p_t$ leads to a specific non-analytic term in the length of the geodesic as a function of boundary time.

We consider radial geodesics that go from one AdS boundary to the other. The geodesics fall towards larger $r$ before bouncing back to the boundary from a maximal $r_\star$, which is the point at which the geodesic comes closest to the singularity. We are interested in 
`symmetric' geodesics that reach $r_\star$ at the `middle' of the extended Penrose diagram so that $\text{Re}\, t(r_\star) = 0$ (the Schwarzschild time coordinate becomes complex beyond the horizon, see \cite{Fidkowski:2003nf}).

Radial spacelike geodesics obey $g_{tt} \dot t^2 + g_{rr} \dot r^2 = 1$. They are characterized by a conserved `energy' $E = - g_{tt} \dot t$. The turning point occurs when $g_{tt}(r_\star) = E^2$ (recall that $g_{tt} > 0$ in the interior). The boundary time for a geodesic with energy $E$ to reach its turning point is
\begin{align}
t(r_\star) - t(0) & = - \int_0^{r_\star} \frac{\sqrt{- g_{tt}g_{rr}}}{g_{tt}} \frac{E dr}{\sqrt{E^2 - g_{tt}}}  \label{eq:t1} \\
& = P \int_0^{r_\star} \frac{\text{sgn}(E) \, e^{\chi/2} dr}{\displaystyle f \sqrt{1 + f e^{-\chi}/(r E)^2}} + \frac{i}{4 T} \,. \label{eq:tg}
\end{align}
In the second step we separated out the imaginary contribution from the pole at $f(r_+) = 0$, with $P$ denoting the principal value. We used the expression (\ref{eq:temp}) for the temperature. Note that the imaginary part is independent of the energy. Because $\text{Re} \, t(r_\star) = 0$ for a symmetric geodesic, while the boundary time $t(0)$ is real, we have that
\be\label{eq:t0}
t(0) = - P \int_0^{r_\star} \frac{\text{sgn}(E) \, e^{\chi/2} dr}{\displaystyle f \sqrt{1 + f e^{-\chi}/(r E)^2}} \,.
\ee

The objective is to find the length $L$ of the geodesic in terms of the boundary time $t(0)$. The regulated length of the geodesic is given in terms of $E$ by
\begin{align}
L & = 2 \int_{r_\text{c}}^{r_\star} \frac{\sqrt{- g_{rr} g_{tt}} dr}{\sqrt{E^2-g_{tt}}} + 2 \log r_\text{c} \\
& = \frac{2}{|E|}\int_{r_\text{c}}^{r_\star} \frac{e^{-\chi/2} dr}{\displaystyle r^2 \sqrt{1 + f e^{-\chi}/(r E)^2}} + 2 \log r_\text{c} \,. \label{eq:L}
\end{align}
We have included an IR regulator $r_\text{c} \to 0$ near the AdS boundary.

The turning point $r_\star$ comes close to the singularity in the large $E$ limit (where the geodesic is almost null). Therefore, we wish to expand both (\ref{eq:t0}) and (\ref{eq:L}) at large $E$. Eliminating $E$ from these expansions will give $L[t(0)]$. Away from the endpoints of the integrals, the integrands of (\ref{eq:t0}) and (\ref{eq:L}) can be expanded in small $1/E^2$. However, non-analytic in $1/E^2$ contributions can arise from the endpoints. This is because $r \to r_\star$ and $r \to r_c \to 0$ do not commute with $E \to \infty$. We consider these limits in turn.

At large $E$, the turning point $r_\star \gg r_+$. The position of the turning point can therefore be evaluated in the asymptotic Kasner form (\ref{eq:cs}) of the metric functions. Therefore
\be\label{eq:rE}
r_\star(E) = \left(\frac{E^2}{f_1 e^{-\chi_1}} \right)^{1/(1-c^2)} + \cdots \quad \text{as} \quad E \to \infty \,.
\ee
The leading non-analytic contribution at large $E$ from the near-singularity endpoint of the integration can be obtained by explicitly performing the integrals in the Kasner scaling regime (in terms of hypergeometric functions). The non-analytic contributions from the near-boundary endpoint can be obtained by
performing the integral in the near-boundary regime. Once the near-boundary expansion (\ref{eq:nearb}) is used in the integrand, the integration variable can be rescaled to $x = r |E|$ and then the integrand can be expanded in $1/E$ and the integration performed order by order.
Putting everything together, the following expansions are found as $E \to + \infty$ (for large negative $E$ note that $L(-E) = L(E)$ while $t(0)(-E) = - t(0)(E)$) 
\begin{align}
L & = 2 \log \frac{2}{E} + \frac{\ell_1}{E} + \frac{\phi_o^2}{4} \frac{1}{E^2} + \frac{\ell_3}{E^3} + \frac{\phi_o\langle \ocal \rangle + 3\braket{T_{tt}}}{3} \frac{\log E}{E^3} + \frac{\ell'_3}{E^{(3+c^2)/(1-c^2)}} + \cdots \,, \label{eq:LL} \\
t(0) & = t_\text{sing} + \frac{1}{E} + \frac{t_2}{E^2} + \frac{\phi_o^2}{12} \frac{1}{E^3} + \frac{t_4}{E^4} + \frac{\phi_o\langle \ocal \rangle + 3\braket{T_{tt}}}{8} \frac{\log E}{E^4} + \frac{t'_4}{E^{4/(1-c^2)}} + \cdots \,. \label{eq:tt}
\end{align}
The coefficients of the terms originating from non-analyticities in the $1/E^2$ expansion are given purely in terms of near-singularity or near-boundary data. These values have been included in the expansions above, and in addition
\begin{align}
\ell'_3 = \sqrt{\pi} (p_t-1) \frac{e^{\chi_1/2 p_t}}{f_1^{(1+p_t)/2p_t}} \frac{\Gamma\left(1/2 + 1/2p_t \right)}{\Gamma\left(1/2p_t\right)} = 2(1-p_t) t'_4 \,.
\end{align}
In contrast, $\ell_1, \ell_3, t_\text{sing}, t_2, t_4, \ldots$ in the expansions above depend on the behavior of the metric along the entire flow. In this sense they are non-universal.

Combining (\ref{eq:LL}) and (\ref{eq:tt}) we obtain, setting $\Delta t = |t(0)- t_\text{sing}|$,
\begin{align}
L & = 2 \log (2 \Delta t) + c_1 \Delta t + c_2 (\Delta t)^2 + c_3 (\Delta t)^3 + \cdots \nonumber \\
& \quad - \frac{\phi_o\langle \ocal \rangle + 3\braket{T_{tt}}}{12} (\Delta t)^3 \log \Delta t + \frac{\sqrt{\pi} p_t e^{\chi_1/2 p_t}}{f_1^{(1+p_t)/2p_t}} \frac{\Gamma\left(1/2 + 1/2p_t \right)}{\Gamma\left( 1/2p_t\right)} (\Delta t)^{-1/p_t}   + \cdots \,. \label{eq:Lt}
\end{align}
The coefficients $c_1,c_2,c_3, \ldots$ are non-universal (depending upon the entire flow). The coefficients of the non-analytic terms in $L(\Delta t)$ are universal and are shown in the above expression.
Recall again that $p_t < 0$. For the Schwarzschild singularity $p_t = -1/3$ and hence the $(\Delta t)^{-1/p_t}$ correction is analytic in that case and cannot be distinguished from the non-universal terms. More generally, however, equation (\ref{eq:Lt}) shows that the scaling exponents of the Kasner singularity determine specific non-analytic corrections to the divergence of the regularized geodesic length as $\Delta t \to 0$. Both the numerical prefactor and the power of these corrections are determined by the near-singularity geometry.

The $(\Delta t)^3 \log \Delta t$ non-analyticity in (\ref{eq:Lt}) originates from the near-boundary region. In general, such terms will be present with non-analytic powers determined by the UV scaling dimensions of the fields (such as the scalar field $\phi$) in the background solution. All scaling dimensions are integers in our case, which is why this non-analyticity is logarithmic. These terms are clearly distinct in origin from the near-singularity terms. In principle the UV non-analyticities are known and can be subtracted out, although this may be delicate in practice with e.g. numerical results.

\subsection{Non-analyticities in finite mass corrections}

The geodesic results of the previous section determine the correlation functions of a large mass scalar field. In this section we show that finite mass corrections to the geodesic result also contain a non-analytic signature of the Kasner exponent. This is because the geometric optics description of the large mass wave equation breaks down in the vicinity of the singularity, where curvature length scales become very small \cite{Fidkowski:2003nf,Festuccia:2005pi}. The result of this section --- which contains a somewhat technical derivation --- is equation (\ref{eq:theanswer}). This shows that finite mass corrections become important when $M (\Delta t)^{-1/p_t} \lesssim 1$. Here $M$ is the mass of the probe scalar field. Recall that $p_t<0$.

The most explicit and direct way to compute finite mass corrections is to find the Green's function for a probe massive scalar field in a systematic large mass WKB expansion.
The Green's function obeys $-\nabla_x^2 G(x,x') + M^2 G(x,x') = \frac{1}{\sqrt{-g}} \delta(x-x')$.
This is the bulk Green's function for a probe scalar field $\psi$ of mass $M$, unrelated to the scalar field $\phi$ in the background. It will be simplest to smear the source over the boundary spatial directions, i.e. take a spatially homogeneous boundary source, so that $\delta(x-x') = \delta(t-t')\delta(r-r')$ in the Green's function equation. In this case, we may look for a solution of the form
\be\label{eq:G1}
G(t,r,t',r') = \int d\omega \hat G(\omega, r, r') e^{i M \omega (t-t')} \,.
\ee
From (\ref{eq:G1}) we have that $\hat G(\omega,r,r')$ obeys
\be
- \frac{d}{dr} \left(\frac{f e^{-\chi/2}}{r^2} \frac{d\hat G}{dr}\right) + M^2 \left( \frac{e^{-\chi/2}}{r^4} - \frac{\omega^2 e^{\chi/2}}{f r^2} \right) \hat G = M \delta(r-r') \,.
\ee
The solution to this equation is (with the location $r'$ of the source outside the horizon, and to start with we also consider $r$ outside the horizon)
\be
\hat G(\omega,r,r') = 
\left\{
\begin{array}{cc}
\displaystyle \frac{\psi_\text{b}(r) \psi_\text{h}(r')}{W} & \qquad r < r' \\
\displaystyle \frac{\psi_\text{h}(r) \psi_\text{b}(r')}{W} & \qquad r > r'
\end{array}
\right. \,.
\ee
Here $\psi_\text{b}$ is the solution to the wave equation (without the delta function source) that is regular at the boundary as $r \to 0$ and $\psi_\text{h}$ is regular (`infalling') at the future horizon as $r \to r_+$. The Wronskian
\be
W = \frac{f e^{-\chi/2}}{M r^2} \left(\psi_\text{h}(r) \psi_\text{b}'(r) -  \psi_\text{h}'(r) \psi_\text{b}(r) \right) \,,
\ee
is independent of $r$.

We are going to be interested in a saddle point of the integral in (\ref{eq:G1}) where $\omega = - i E$, with $E$ real. This will correspond to a particular spacelike geodesic in the WKB limit. We will therefore take this imaginary value for $\omega$ in the following. In the remainder we will work with $E>0$ (for concreteness).
Within a large $M$ WKB expansion we have
\be
\psi_\text{b}(r) = F(r) \exp\left\{M \int^r_{r_\text{c}} \left[ s_0(x) + \frac{s_2(x)}{M^2} + \cdots \right] \, dx \right\} \,,
\ee
with $r_\text{c}$ a regulator close to the boundary. Our objective is to compute the leading corrections away from the large mass limit, hence we keep the order $1/M$ term in the exponent. The functions appearing here are
\begin{align}
F(r) & = r \left(E^2+\frac{f e^{-\chi}}{r^2}\right)^{-1/4} \,, \\
s_0(r) & = \frac{e^{\chi/2}}{f} \left( E^2 + \frac{f e^{-\chi}}{r^2}\right)^{1/2} \,, \\
s_2(r) & = \frac{1}{2 s^2_0} \frac{(F')^2}{F^3} \left(\frac{s_0 F^2}{F'} \right)'  \,.\label{eq:s22}
\end{align}
This solution decays as $r \to 0$ (with $r_c$ fixed). The other solution is
\be
\psi_\text{h}(r) = F(r) \exp\left\{M \int^{r_+}_r \left[ s_0(x) + \frac{s_2(x)}{M^2} + \cdots \right] \, dx \right\} \,. \label{eq:infal}
\ee
As previously, $r_+$ is the horizon. The solution (\ref{eq:infal}) has $\psi_\text{h}(r) e^{M E t}$ regular on the future horizon, which is where $E>0$ geodesics cross the Einstein-Rosen bridge. With these solutions, the Wronskian is
\be
W = 2 \exp\left\{M \int^{r_+}_{r_c} \left[ s_0(x) + \frac{s_2(x)}{M^2} + \cdots \right] \, dx \right\} \,.
\ee
This is indeed manifestly a constant.
The prefactor of the exponential is given to the order we are working throughout.

The Green's function is then
\be\label{eq:G2}
\hat G(-iE,r,r') = \half F(r) F(r') \exp\left\{-M \int^{r}_{r'} \left[ s_0(x) + \frac{s_2(x)}{M^2} + \cdots \right] \, dx\right\} \,, \qquad \text{for } r > r'.
\ee
And for $r < r'$ the order of limits of the integration in (\ref{eq:G2}) is reversed. At this point we have both $r$ and $r'$ outside the horizon. At large $M$ the integral over $E$ in (\ref{eq:G1}) can be evaluated by saddle point. The saddle point is at
\be
t - t' = \int_{r'}^r \frac{e^{\chi/2} dr}{f \sqrt{1 + f e^{-\chi}/(r E)^2}} \,. \label{eq:tE}
\ee
This agrees with the geodesic relation (\ref{eq:tg}).
Equation (\ref{eq:tE}) should be read as specifying $E$ as a function of $t-t'$, given $r$ and $r'$. A similar analysis was done previously in \cite{Festuccia:2005pi}.

We now wish to analytically continue (\ref{eq:G2}) past the horizon. We take the source to be at $r' = r_c$ and take $r$ past the horizon. As explained in e.g. \cite{Fidkowski:2003nf, Festuccia:2005pi}, we can do this at the cost of incurring imaginary shifts in the time difference in (\ref{eq:tE}). We already saw this shift in (\ref{eq:tg}). Furthermore, we wish to take the geodesic down to the turning point $r_\star$ and then out to the boundary on the other side of the Einstein-Rosen bridge. However, the WKB form (\ref{eq:G2}) is not valid very close to the turning point. As usual, this is manifested in divergences in $F(r)$ and the integral of $s_2(r)$ as we approach the turning point. This can be dealt with by deforming the contour of integration into the complex $r$ plane close to $r_\star$, so that it encircles the turning point at some small radius $\vep$, while remaining within the domain of validity of the WKB approximation \cite{PhysRev.41.713}. The full contour therefore goes from $r_c$ to $r_\star - \vep$, loops around $r_\star$ in the complex $r$ plane, and then runs from $r_\star - \vep$ back to $r_c$. Because of the square root branch point at $r=r_\star$, the integrals from $r_c$ to $r_\star - \epsilon$ and back add rather than cancel. The contribution from the loop around $r_\star$ precisely cancels out any divergent terms at the turning point as $\vep \to 0$. Thus we obtain the boundary-to-boundary Green's function
\be\label{eq:G3}
\hat G_{12}(-iE) = \half F(r_c)^2 \exp\left\{-2 M \int^{r_\star - \vep}_{r_c} \left[ s_0(x) + \frac{s_2(x)}{M^2} + \cdots \right] \, dx\right\} \,,
\ee
where we throw away any terms in the exponent that diverge as $\vep \to 0$.

Performing the integral over $E$ in (\ref{eq:G1}) by saddle point gives
\begin{align}
G_{12}(t') & = \half F(r_c)^2 D(r_c,r_\star) \exp\left\{- M L - \frac{2}{M} \int_{r_c}^{r_\star - \vep} s_2(x) dx + \cdots\right\} \,. \label{eq:Gfinal}
\end{align}
Here the (unregulated near $r = 0$) geodesic length
\be
L = 2 \int_{r_c}^{r_\star} \frac{e^{-\chi/2} dr}{r^2 \sqrt{E^2 + f e^{-\chi}/r^2}} \,.
\ee
This is of course exactly the factor that appeared previously in (\ref{eq:L}). Restricting to symmetric geodesics leaving the right boundary at time $t'$,
then the time at the left boundary $t = -t' + i/2T$ in (\ref{eq:tE}) and hence $E = E(t')$ is given by
\be
t' = - P \int_{r_c}^{r_\star} \frac{e^{\chi/2} dr}{f \sqrt{1 + f e^{-\chi}/(r E)^2}} \,. 
\ee
The remaining $D(r_c,r_\star)$ term in (\ref{eq:Gfinal}) is due to fluctuations about the saddle, and we will evaluate it shortly.

We now want to evaluate (\ref{eq:Gfinal}) as $\Delta t = t' - t_\text{sing} \to 0$. From (\ref{eq:tt}) we know that in this regime $E \approx 1/\Delta t \to \infty$. From (\ref{eq:Lt}) we know that $L \approx 2 \log (2 \Delta t/r_c)$. The $r_c$ appears here because the $L$ in (\ref{eq:Lt}) is regulated. The integral $\int s_2(x) dx$ in the exponent of (\ref{eq:Gfinal}) has three contributions at large $E$: from near the turning point, from the intermediate region, and from near the boundary. Recall that $s_2(x)$ was given in (\ref{eq:s22}). In the intermediate region the integrand can be expanded in $E$, and gives a leading contribution of order $1/E$. Near the boundary there is a leading contribution of $\frac{9}{8} \log [2/(E r_c)]+\frac{1}{12} \approx \frac{9}{8} \log(2 \Delta t/r_c) + \frac{1}{12}$. Near the singularity the integral can be performed in terms of hypergeometric functions and one obtains (after subtracting off diverging terms as $\epsilon \to 0$, as discussed above) the contribution $\sqrt{\pi f_1} r_\star^{(3+c^2)/2} \Gamma(1/2 - 1/2 p_t)/[6 \Gamma(-1/2p_t)]$. Here we used (\ref{eq:rE}) to relate $E$ and $r_\star$. This singular term from the near-singularity region diverges as $r_\star \to \infty$ and therefore dominates the integral. The singular logarithmic term from the boundary is also important, it will describe a $1/M^2$ correction to the CFT anomalous dimension of the heavy operator.

We must furthermore include fluctuations about the saddle of the $E$ integral. These contribute several terms at the same order to the terms we have been considering. Let $S_0 = \int s_o(x) dx$ and $S_2 = \int s_2(x) dx$, with both integrals from $r_c$ to $r_\star - \vep$. As previously, divergences are to be removed as $\vep \to 0$. Performing the Gaussian integral about the saddle point and keeping all terms that contribute to order $1/M$ we obtain the fluctuation contribution
\be
D = \sqrt{\frac{\pi}{M \pa^2_E S_0}} \left[1 + \frac{1}{M}\left(\frac{\pa_E^2 F^2}{4 F^2} \frac{1}{\pa^2_E S_0} - \frac{\pa_E F^2}{4 F^2} \frac{\pa^3_E S_0}{(\pa^2_E S_0)^2} -  \frac{1}{16} \frac{\pa^4_E S_0}{(\pa^2_E S_0)^2} + \frac{5}{48} \frac{(\pa^3_E S_0)^2}{(\pa^2_E S_0)^3} \right) + \cdots \right] \,. \label{eq:D}
\ee
Thus we need to consider the integral
\begin{align}
\pa_E^2 S_0 = \int_{r_c}^{r_\star-\vep} \frac{e^{-\chi/2} dr}{r^2 \left(E^2 + f e^{-\chi}/r^2 \right)^{3/2}} \,, \label{eq:sum}
\end{align}
with the higher derivatives $\pa_E^3 S_0$ and $\pa_E^4 S_0$ obtained by differentiating. Due to the regularization of the endpoint of the integral, in which divergent terms are subtracted out, there is no contribution from when the derivative hits the upper limit $r_\star(E)$. Note furthermore that as $r_c \to 0$ (the cutoff should be above the energy scale $E$), then $\pa_E^2 F^2/F^2 \sim E^0$ and $\pa_E F^2/F^2 \sim E$. The integral in (\ref{eq:sum}) has contributions from the near-singularity regime, from the intermediate regime and from the near-boundary regime. These contributions are, respectively, as follows: $r_\star^{(c^2-5)/2};1/E^3;1/E^2$. The near-boundary contribution dominates, so that $\pa_E^2 S_0 \approx 1/E^2$ at large $E$ (with the given prefactor). It follows that $\pa_E^3 S_0 \approx 1/E^3$ and $\pa_E^4 S_0 \approx 1/E^4$. Using these scalings in (\ref{eq:D}) we see that the largest of the terms in the round brackets goes like $E^2$ at large $E$. This is subleading at large $E$ compared to the non-fluctuation $r_\star^{(3+c^2)/2} \sim E^{(3+c^2)/(1-c^2)}$ contribution that we found from $S_2$ above. Therefore to leading order at large $E$ we have simply
\be
D \approx \sqrt{\frac{\pi}{M \pa^2_E S_0}} \approx E \sqrt{ \frac{\pi}{M}} \,.
\ee

Putting everything together in (\ref{eq:Gfinal}), and using the expression (\ref{eq:rE}) for $r_\star$ along with $E \approx 1/(\Delta t)$, we finally obtain, to order $1/M$,
\be\label{eq:theanswer}
G_{12}(t') = \sqrt{ \frac{\pi}{M}} (2 \Delta t)^2 \left(\frac{r_c}{2 \Delta t} \right)^{2(M + \frac{3}{2} + \frac{9}{8M})} \left(1 + \frac{g_1}{M (\Delta t)^{-1/p_t}} + \cdots \right)  \,.
\ee
The coefficient $g_1$ depends only on near-singularity data:
\be
g_1 = \frac{\sqrt{\pi}}{6} \frac{f_1^{(1+p_t)/2p_t}}{e^{\chi_1/2 p_t}} \frac{\Gamma(1/2 - 1/2 p_t)}{\Gamma(-1/2p_t)} \,.
\ee
In (\ref{eq:theanswer}) we have grouped terms in the exponent that correspond to the large $M$ expansion of the UV scaling dimension of the operator: $\Delta = \frac{1}{2} \left(3 + \sqrt{9 + 4 M^2}\right) = M + \frac{3}{2} + \frac{9}{8 M} + \cdots$. The correlator of the dual field theory is obtained from the bulk result (\ref{eq:theanswer}) by stripping off the conformal factor of $r_c^{2 \Delta}$. The extra overall factor of $(2 \Delta t)^2$ in (\ref{eq:theanswer}) is present because we have computed the Green's function for a source that is homogeneous in space.
Indeed, precisely this factor is expected in CFT correlation functions $G(t,\vec k = 0)$.

The result (\ref{eq:theanswer}) for the correction agrees with an estimate that can be made as in \cite{Fidkowski:2003nf}. At large $M$ the contribution to the Green's function from near the singularity can be estimated from a short time heat kernel expansion as
\be
\delta G \sim \frac{R^2 \delta \tau}{M} \,.
\ee
Here $\delta \tau$ is the proper time the geodesic spends close to the singularity and $R$ is the curvature scale (with units of inverse length) experienced by the geodesic in that time. Let $\tau_\star$ be the closest proper distance the geodesic comes to the singularity. Then, because this is only scale in the near-singularity region, $R \sim 1/\tau_\star$ and $\delta \tau \sim \tau_\star$. Recall from below (\ref{eq:cs}) that $\tau_\star \sim r_\star^{-(3+c^2)/2} \sim (\Delta \tau)^{-1/p_t}$. Thus $\delta G \sim 1/[M (\Delta \tau)^{-1/p_t}]$, in agreement with (\ref{eq:theanswer}).

The most important part of (\ref{eq:theanswer}) for our purposes is the $1/M$ correction. Recall that $p_t$ was given in (\ref{eq:ppp}) and is negative. The correction therefore shows that the large $M$ limit does not commute with the near-singularity $\Delta t \to 0$ limit. The crossover regime between these two limits is determined by the Kasner exponent as $M (\Delta t)^{-1/p_t} \sim 1$. It may be interesting in the future to probe this crossover even more explicitly by solving the finite $M$ wave equation in the bulk numerically. As in the previous section, recall that $p_t = -1/3$ for Schwarzschild, so that the correction is given by an integer powers of $\Delta t$ in that case, but not in general.

\subsection{Geodesics at all energies}

It was explained in \cite{Fidkowski:2003nf} that the divergence of $L$ as $t \to t_\text{sing}$ cannot occur on the `physical' sheet of the correlation function but can only be accessed by analytic continuation in time. In this section we shall see that the same conclusion holds for the geometries we are looking at, with one difference.

\begin{figure}[ht!]
    \vskip 0.1in
    \centering
    \includegraphics[width=0.75\textwidth]{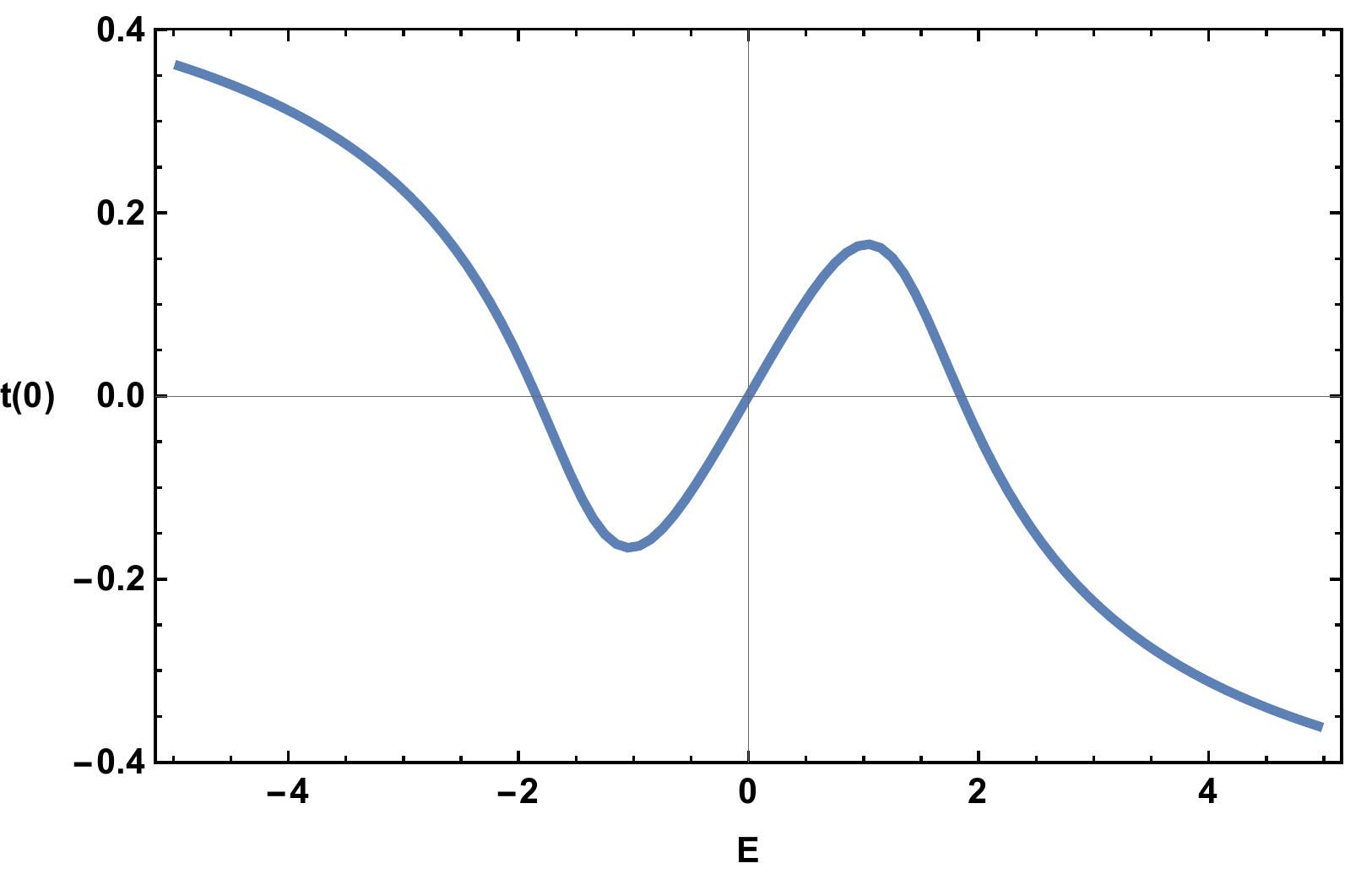}
    \caption{Boundary time versus geodesic energy for the black hole geometry with $\phi_o/T \approx 7$ and cutoff $r_c = 0$. The shape of the curve --- including the $t(0) \propto E$ behavior close to $E=0$ --- is the same for AdS-Schwarzschild in 3+1 bulk dimensions.}
    \label{fig:tE}
    \vskip -0.1in
\end{figure}

Figure \ref{fig:tE} shows $t(0)$ as a function of $E$. Previously we have focused on large $E$.
For small $t(0)$ there are three allowed values of $E$.
The point made in \cite{Fidkowski:2003nf} is that the correlation function defined by continuation from Euclidean space chooses the branch for which $t(0) \propto E$ as $E \to 0$. This is not the branch which connects to the large $E$ regime that we have studied. Instead, the physical branch moves into the complex $E$ plane for $|t(0)|$ greater than the extrema seen in Fig. \ref{fig:tE}. These will correspond to geodesics in the complexified $r$ coordinate that do not probe the singularity (see further comments at the end of the following section, these complex saddles describe the physical quasinormal modes of the black hole).

A difference with the results in \cite{Fidkowski:2003nf} is that even with the cutoff $r_c = 0$, the two extrema in Fig. \ref{fig:tE} are not degenerate. This is not due to the scalar field in our solutions, it is also the case for the Schwarzschild-AdS background in 3+1 bulk dimensions. At small $E$, therefore, $t(0) \sim E$ while the length $L \sim \text{const.} + E^2$, so that $L \sim \text{const.} + t(0)^2$, and there is no branch point in $L[t(0)]$ at $t(0) = 0$.

\subsection{Entanglement entropy probe}

Entanglement entropy is obtained in holographic models as the area of extremal surfaces in the bulk geometry \cite{Ryu:2006bv, Hubeny:2007xt}. The entangled spatial subregion of the CFT can be taken to be the same region --- for example, half of space --- on both copies of the thermofield double state. In this case the extremal surface straddles the Einstein-Rosen bridge in a way analogous to the geodesics we have considered thus far \cite{Hartman:2013qma}.
There is, however, an important difference between the extremal surfaces and the geodesics. At late times, the extremal surface gets stuck on a specific critical constant-$r$ surface inside the horizon and does not approach the singularity. The surface fills out the critical surface at late times, leading to a linear growth in entanglement entropy with time. This growth defines a velocity that is sensitive to the black hole interior, but not to the near-singularity region. We will briefly review these facts, applied to our geometries.

The bulk surface is extended in the boundary $y$ direction, fixed at boundary $x = x_o$ and follows a curve $r(t)$ in the bulk. Calculations that are very analogous to the geodesic case show that a symmetric surface at boundary time $t(0)$ reaches a radius $\hat r_\star$ with
\be\label{eq:t00}
t(0) = - P \int_{0}^{\hat r_\star} \frac{\text{sgn}(\hat E) e^{\chi/2} dr}{f \sqrt{1 + f e^{-\chi}/(r^2 \hat E)^2}} \,.
\ee
This expression is almost identical to the geodesic result (\ref{eq:t0}), but there is an extra factor of $r^2$ in the square root. This is because the `energy' $\hat E$ must now take into account the extension of the surface along the boundary $y$ direction, with $g_{yy} = 1/r^2$. This extra factor makes an important difference because as $r \to \infty$ in the Kasner regime then
\be
\frac{f e^{-\chi}}{r^2} \sim r^{1-c^2} \to \infty \,, \qquad \text{but} \qquad \frac{f e^{-\chi}}{r^4} \sim r^{-1-c^2} \to 0 \,. 
\ee
It is also clear that $f e^{-\chi}/r^4 = 0$ on the horizon. 
It follows that $-f e^{-\chi}/r^4$ has a maximum at some radius $r_\text{crit}$ inside the horizon. When $\hat E$ is such that $1 + f e^{-\chi}/(r^2 \hat E)^2 = 0$ at $r = r_\text{crit}$, then the function in the square root in (\ref{eq:t00}) has a double zero at $r = r_\text{crit}$ and the integral diverges. That is, $t(0) \to \infty$ and $\hat r_\star = r_\text{crit}$. Thus at late times the surfaces do not reach a turning point close to the singularity, as was the case for geodesics, but rather get stuck at the critical radius. See \cite{Hartman:2013qma} for a more extended discussion. General obstructions to extremal surfaces reaching spacelike singularities are discussed in \cite{Wall:2012uf, Engelhardt:2013tra}.

\begin{figure}[ht!]
    \vskip 0.1in
    \centering
    \includegraphics[width=0.75\textwidth]{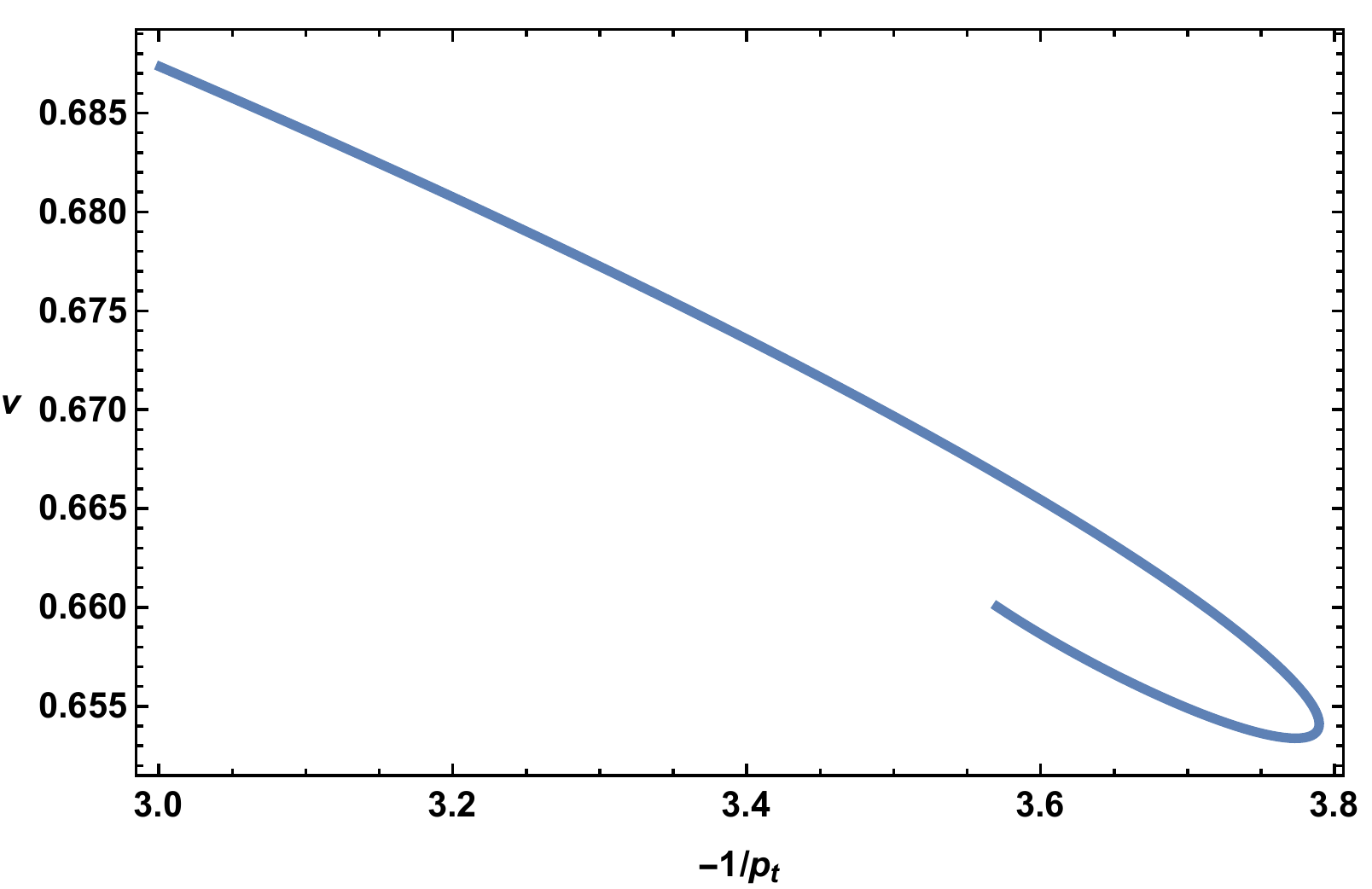}
    \caption{Entanglement velocity as a function of $-1/p_t$. The velocity decreases away from the Schwarzschild value of $v = \sqrt{3}/2^{4/3}$ at $-1/p_t = 3$. Geometries from Fig. \ref{fig:kasner} are plotted.}
    \label{fig:ent}
    \vskip -0.1in
\end{figure}

It is easily seen that at late times the entanglement entropy $S$, given by the area of the extremal surface, grows linearly with time \cite{Hartman:2013qma}. Essentially, the surface grows along the $r=r_\text{crit}$ slice. This late time linear growth defines an entanglement velocity $v$ according to
\be
\frac{dS}{dt(0)} = v V_1 s \,, \qquad v^2 = r_+^4 \left. \frac{|f| e^{-\chi} }{r^4}\right|_{r=r_\text{crit}} \,.
\ee
Here $V_1$ is the length of the $y$ boundary direction and $s$ is the thermal entropy density. Fig. \ref{fig:ent} shows the entanglement velocity for our solutions with a deformed interior. While these velocities do not probe the near-singularity region, we have plotted them as a function of the Kasner exponent to emphasize that the velocity is a property of the black hole interior.

As discussed in \cite{Hartman:2013qma}, boundary to boundary geodesics in fact also exhibit critical radii leading to a linear dependence of the geodesic length with time. These radii are typically complex, and the corresponding time dependence describes the quasinormal ringdown of the black hole (see also \cite{Motl:2003cd}). These complex saddles are precisely the physical saddles mentioned in the previous section, that determine the late time behavior of correlation functions.

\section{Discussion}

A major motivation for probing behind black hole horizons has been to capture quantum gravitational phenomena in the vicinity of the singularity (or elsewhere). However, classical gravity is also expected to be extremely rich in the approach to the singular region. This rich dynamics should be contained within the thermal state of a dual large $N$ CFT. In this work we have demonstrated a simple instance of how this can work. We have firstly shown that deformation of the thermal CFT state by a relevant operator leads to a deformation of the Schwarzschild singularity, at late interior times, to a more general Kasner universe, as in Fig. \ref{fig:kasnerflow}. The Kasner region is characterized by the Kasner exponents. Secondly, we have shown that these exponents universally (in the sense that only near-singularity data is needed) determine non-analytic corrections to correlation functions of the dual thermal CFT. These correlation functions must be analytically continued into a near-singularity regime, described by a spacelike geodesic that crosses the Einstein-Rosen bridge and comes close to the singularity \cite{Fidkowski:2003nf,Festuccia:2005pi}. A curious aspect of the non-analytic corrections we have discussed is that they become analytic for the (non-generic) Schwarzschild singularity, where $p_t = -1/3$ leads to integer powers.

The interior of charged black holes has been the subject of extensive research due to the presence of Cauchy horizons. Uniform deformations of the boundary theory, of the kind we have investigated here, may give a simple holographic laboratory for those questions.

In this work we have restricted to deformations and probes that are uniform in boundary space and time. More generically, the classical approach to the singularity is expected to be highly inhomogeneous. A natural observable to capture this more general dynamics at a perturbative level is the OTOC, in which a localized boundary perturbation is probed at later times \cite{Shenker:2013pqa, Roberts:2014isa}. For our purposes, the later probe would access the instability of the Schwarzschild near-singularity geometry due to the initial perturbation, rather than the sensitivity to initial conditions caused by high energy near-horizon processes. We hope to report on results in this direction in the future.

\section*{Acknowledgements}

It is a pleasure to thank Gary Horowitz and Stephen Shenker for comments on a draft. ZS would like to thank Jonathan Luk for helpful discussions on black hole instabilities. JK is supported by the Simons Foundation. SAH is partially supported by DOE award de-sc0018134 and by a Simons Investigator award.

\bibliographystyle{ourbst}
\bibliography{rgflow}

\end{document}